\documentclass[twocolumn,tighten,times]{aastex61}
\usepackage[utf8]{inputenc}
\usepackage{amssymb}
\usepackage{amsmath}
\usepackage{amsfonts}
\usepackage{booktabs}
\usepackage{tabularx}

\def\be{\begin{equation}}
\def\ee{\end{equation}}
\def\ba{\begin{eqnarray}}
\def\ea{\end{eqnarray}}

\newcommand{\DM}{\mathrm{DM}}
\newcommand{\IGM}{\mathrm{IGM}}
\newcommand{\ISM}{\mathrm{ISM}}
\newcommand{\CGM}{\mathrm{CGM}}
\newcommand{\host}{\mathrm{host}}

\newcommand{\pcc}{pc\,cm$^{-3}$}
\newcommand{\pccc}{pc\,cm$^{-6}$}

\newcommand{\EM}{\mathrm{EM}}

\begin{document}

\title{Estimates of Fast Radio Burst Dispersion Measures from Cosmological Simulations}

\author[0000-0002-8826-1285]{N.\, Pol}
\affil{Department of Physics and Astronomy, West Virginia University, White Hall, Morgantown, WV 26506, USA; nspol@mix.wvu.edu}
\affil{Center for Gravitational Waves and Cosmology, West Virginia University, Chestnut Ridge Research Building, Morgantown, WV 26505}
\author[0000-0003-0721-651X]{M.\,T.\,Lam}
\affil{Department of Physics and Astronomy, West Virginia University, P.O. Box 6315, Morgantown, WV 26506, USA}
\affil{Center for Gravitational Waves and Cosmology, West Virginia University, Chestnut Ridge Research Building, Morgantown, WV 26505, USA}
\author[0000-0001-7697-7422]{M.\,A.\,McLaughlin}
\affil{Department of Physics and Astronomy, West Virginia University, P.O. Box 6315, Morgantown, WV 26506, USA}
\affil{Center for Gravitational Waves and Cosmology, West Virginia University, Chestnut Ridge Research Building, Morgantown, WV 26505, USA}
\author{T.\,J.\,W.\,Lazio}
\affil{Jet Propulsion Laboratory, California Institute of Technology, 4800 Oak Grove Drive, Pasadena, CA 91109, USA}
\author[0000-0002-4049-1882]{J.\,M.\,Cordes}
\affil{Department of Astronomy and Cornell Center for Astrophysics and Planetary Science, Cornell University, Ithaca, NY 14853, USA}

\begin{abstract}
    We calculate the dispersion measure (DM) contributed by the intergalactic medium (IGM) to the total measured DM for fast radio bursts (FRBs). We use the MareNostrum Instituto de Ciencias del Espacio (MICE) Onion Universe simulation \citep{mice_onion_universe} to track the evolution of the dark matter particle density over a large range of redshifts. We convert this dark matter particle number density to the corresponding free electron density and then integrate it to find the DM as a function of redshift. This approach yields an intergalactic DM of $\DM_{\IGM}(z = 1) = 800^{+7000}_{-170}$~\pcc, with the large errors representative of the structure in the IGM. We place limits on the redshifts of the current population of observed FRBs. We also use our results to estimate the host galaxy contribution to the DM for the first repeater, FRB 121102, and show that the most probable host galaxy DM contribution, $\DM_{\host} \approx 310$~\pcc, is consistent with the estimate made using the Balmer emission lines in the spectrum of the host galaxy, $\DM_{\rm Balmer} = 324$~\pcc\ \citep{121102_localzn_2} We also compare our predictions for the host galaxy contribution to the DM for the observations of FRB 180924 \citep{askap_loczn} and FRB 190523 \citep{dsa10_loczn}, both of which have been localized to a host galaxy.
\end{abstract}

\keywords{methods: statistical ---
          pulsars: general ---
          galaxies: intergalactic medium}

\section{Introduction}
    
    Fast radio bursts (FRBs) are extragalactic transient radio sources which emit bright ($\sim$Jy), millisecond duration bursts. The first FRB was discovered by \citet[][]{lorimer_burst} in archival Parkes Telescope data. Since then, at least an additional 64 FRBs have been detected with different telescopes around the world \citep{frbcat}. Of these 65 FRBs, only two, FRB 121102 \citep{121102_disc, 121102_repeats} and FRB 180814.J0422+73 \citep{second_repeater_disc}, are known to have emitted multiple bursts. Out of these two, FRB 121102 has been localized \citep{121102_localzn_1} to a dwarf galaxy at a redshift of $z = 0.197$ \citep{121102_localzn_2} with stellar mass $\sim$$10^8$~M$_{\odot}$ \citep{Bassa_121102_stellar_mass}. Recently, two more FRBs have been localized in redshift and to their host galaxies using interferometric telescopes. \citet[][]{askap_loczn} used the Australian Square Kilometer Array Pathfinder (ASKAP) telescope to localize FRB 180924 to a redshift of $z = 0.32$ and an early-type spiral galaxy with a total stellar mass of $\sim$$10^{10}$~M$_{\odot}$. Similarly, \citet{dsa10_loczn} used the Deep Synoptic Array (DSA) to localize FRB 190523 to a redshift of $z = 0.66$ and a galaxy with stellar mass $\sim$$10^{11}$~M$_{\odot}$.
    
    When FRBs were first discovered \citep{lorimer_burst, thornton_bursts}, it was observed that these bursts had a dispersion measure (DM\footnote{In this work, we follow the convention from the FRB literature that DM stands for ``dispersion measure'' and not ``dark matter''.}) that was significantly higher than the Milky Way contribution to the DM in that direction calculated using the NE2001 Galactic free electron density model \citep{ne2001}. This led many to believe that these bursts had an extragalactic, if not cosmological, origin. This was confirmed when FRB 121102 was found to be of cosmological origin with the localization and subsequent redshift measurement of $z = 0.197$ \citep{121102_localzn_2}. 
    
    In general, the DM can be calculated by integrating the free electron density, $n_{\rm e}(z)$, along a given line of sight, $dl$, up to a redshift $z_{\rm max}$ \citep{ioka_dm, inoue_dm, deng_frb, mcquinn_dm_igm},
    \begin{equation}
        \displaystyle \DM (z_{\rm max}) = \int_{0}^{z_{\rm max}} \frac{n_{\rm e}(z)}{1 + z} \, dl,
        \label{cosmo_dm_eq}
    \end{equation}
    where,
    \begin{equation*}
        \displaystyle dl = c \, \left| \frac{dt}{dz} \right| \, dz
    \end{equation*}
    and
    \begin{equation*}
        \displaystyle \left| \frac{dt}{dz} \right| = \frac{1}{(1 + z) H(z)}
    \end{equation*}
    At very low redshifts, $z \approx 0$, for example inside the Milky Way, Eq.~\ref{cosmo_dm_eq} reduces to
    \begin{equation}
        \displaystyle \DM = \int_{0}^{L} n_{\rm e}(l) \, dl,
        \label{OG_dm_eq}
    \end{equation}
    where $L$ is the distance in parsecs. Note that the free electron density is in units of cm$^{-3}$ and the distance has units of parsecs (pc), giving DM the units of \pcc.
    The observed DM of an FRB at redshift $z$, $\DM_{\rm obs}$, can be split up into the sum of the contributions of its line-of-sight components
    \be 
    \DM_{\rm obs}(z) = \DM_{\ISM} + \DM_{\CGM} + \DM_{\IGM}(z) + \frac{\DM_{\host}}{1 + z},
    \label{eq:DMmodel}
    \ee
    where $\DM_{\ISM}$ represents the contribution from the interstellar medium (ISM) in the Milky Way, $\DM_{\CGM}$ represents the DM contribution from the circumgalactic medium (or halo) around the Milky Way, $\DM_{\IGM}$ represents the DM contribution from the intergalactic medium (IGM), and $\DM_{\host}$ represents the DM contribution from both the interstellar medium in the host galaxy and the local environment of the FRB. 
    
    The Milky Way contribution, $\DM_{\ISM}$, calculated using Eq.~\ref{OG_dm_eq}, can be modeled
    using the NE2001 Galactic free electron density model \citep{ne2001} or the free electron density model proposed by \citet[][YMW model]{YMW16}. 
    The circumgalactic medium contribution is usually assumed as $\DM_{\CGM} = 30$~\pcc\ \citep{dolag_dm_igm}, though there will be some directional dependence because of our offset position with respect to the Galactic center. A more detailed modeling of the circumgalactic medium including the directional dependence was done by \citet[][]{prochaska_dm_cgm}, who found that a better estimate for the dispersion measure contribution from the CGM was $50 \, {\rm pc\,cm^{-3}} < \DM_{\CGM} < 80$~\pcc. We use the estimate made by \citet[][]{prochaska_dm_cgm} in this analysis.
    This leaves two unknown quantities in Eq.~\ref{eq:DMmodel}, the intergalactic $\DM_{\IGM}$ contribution and the host galaxy $\DM_{\host}$ contribution. 
    
    Since the source (or sources) of FRBs are not yet known (see  \citet{frb_source_theory_cat} (and the associated FRB Theory Wiki\footnote{\url{https://frbtheorycat.org/index.php/Main_Page}}) for a collection of all source models presented to date), it makes theoretically estimating the local environmental contribution to the total FRB DM difficult. In addition, since FRBs are located at large distances and most have not been localized to a host galaxy, it is extremely difficult to directly probe the environment of the FRB through observations. This makes it hard to develop models which will estimate the value or shape of the DM contribution from the host galaxy of the FRB, $\DM_{\host}$.
    
    On the other hand, it is relatively straightforward to theoretically derive the IGM's contribution to the DM using Eq.~\ref{cosmo_dm_eq} and assuming a cosmic reionization history \citep{ioka_dm, inoue_dm, deng_frb, mcquinn_dm_igm}.
    However, it is difficult to try to observationally constrain the intergalactic DM, $\DM_{\IGM}$. The delay in the pulse arrival time due to dispersion is inversely proportional to the observing frequency, $\propto \nu^{-2}$, the effect of which is observable at radio frequencies. Due to a lack of extragalactic transient sources which are bright enough to be visible in the radio band and narrow enough to allow DM measurements, there had been no empirical measurements of intergalactic DMs outside of the Milky Way and Magellanic Clouds until the discovery of FRBs. In spite of the lack of empirically measured values for $\DM_{\IGM}$, we can estimate the IGM's contribution to the total DM measured for FRBs by using Eq.~\ref{cosmo_dm_eq}.
    
    This approach of calculating the intergalactic DM has been performed in quite a few studies, the most popular of which are those by \citet[][]{ioka_dm}, \citet[][]{inoue_dm}, and \citet[][]{mcquinn_dm_igm}. Assuming a current free electron density which evolves as $(1 + z)^3$ and different cosmic reionization histories, \citet[][]{ioka_dm} and \citet[][]{inoue_dm} were able to estimate the intergalactic DM out to redshifts as high as $z = 30$. Another method makes use of large cosmological simulations to simulate the evolution of the intergalactic medium as a function of redshift. This method was first introduced by \citet[][]{dolag_dm_igm}, where they used simulations of the Galactic halo \citep{beck_halo_sim} to estimate the circumgalactic contribution to the DM for FRBs. They also developed and made use of the \textit{Magneticum Pathfinder}\footnote{\url{http://www.magneticum.org/}} \citep[][]{dolag_dm_igm} data set to simulate the dark matter and gas particle particle number density in the universe as a function of redshift. Using these simulations, they were able to produce maps of the free electron density, and thus, the DM as a function of redshift.
    
    The advantages of trying to estimate the intergalactic DM are twofold. One, we can place upper limits on the redshift of the FRBs. If we assume a host galaxy contribution, $\DM_{\host} = 0$, then we can solve Eq.~\ref{eq:DMmodel} for $\DM_{\IGM}$, and then use the derived intergalactic DM--$z$ relation to place an upper limit on the redshift of the FRB. An example of such an $\DM_{\IGM}$--$z$ relation is that derived by \citet[][]{ioka_dm} and \citet[][]{inoue_dm}, which at low redshifts ($z \leq 2$) can be expressed as
    \begin{equation}
        \displaystyle \DM_{\IGM} = 1200\,z \, [{\rm pc \, cm}^{-3}].
        \label{inoue_reln}
    \end{equation}
     Placing such limits on the redshift of the FRBs is useful when performing follow-up studies to locate the host galaxy of the FRB in archival data or estimate the cosmological distribution of the FRBs to, for example, place limits on the baryon mass fraction, $\Omega_{\rm bary}$, or probe the reionization history of the universe \cite[][]{cosmo_w_frb_Deng}. The second advantage of estimating the intergalactic DM is that it allows us to place an estimate on the host galaxy DM distribution. Looking at Eq.~\ref{eq:DMmodel}, we can see that once we know the distribution of $\DM_{\IGM}$, we can directly compute the distribution for $\DM_{\host}$ (within the errors on the estimated Milky Way and halo contributions). This will allow us to probe the region local to the FRB itself. Depending on how constraining the derived $\DM_{\host}$ distribution is, this might even allow us to rule out some of the proposed source models for FRBs.
    
    In this work, we use cosmological simulations made by the MareNostrum Instituto de Ciencias del Espacio (MICE) team \citep{mice_onion_universe} which simulate the evolution of large scale structure through dark matter particles in a redshift range $0 < z < 1.4$. We describe how we convert from the dark matter particle number densities reported in these simulations to baryonic matter density, and then the intergalactic DM in Sec.~\ref{sims}. Next, we develop intergalactic DM probability distributions at different redshifts in Sec.~\ref{stats_from_maps}. Using these distributions, we place upper limits on the redshifts for all the observed FRBs in Sec.~\ref{redshift_secn}. We exploit the localization of FRB 121102 to a host galaxy at a redshift $z = 0.19$ \citep{121102_localzn_2} to place limits on the distribution of the host galaxy DM in Sec.~\ref{host_dm_secn}. We offer our conclusions in Sec.~\ref{conclusion}.
    
    In this work, we assume a flat concordance $\Lambda$CDM \citep{planck_2018} model with the matter density, $\Omega_m = 0.315$, dark energy density, $\Omega_{\Lambda} = 0.685$, baryonic matter density, $\Omega_{\rm bary} h^2 = 0.0224$, and Hubble constant, $H_0 = 100 \, h$~km~s$^{-1}$~Mpc$^{-1}$ with $h = 0.7$.

\section{Simulations} \label{sims}
    
    For this analysis, we make use of the MareNostrum Instituto de Ciencias del Espacio (MICE) Onion Universe\footnote{\url{http://maia.ice.cat/mice/}} simulation \citep[][]{mice_onion_universe}. This is a large N-body dark matter simulation with $2048^3$ dark matter particles in a box-size of 3072~Mpc/$h$. The mass of the dark matter particles in this simulation is $M_{\rm dark} = 2.9 \times 10^{10}$~M$_{\odot}$/$h$. The output of the simulation is provided in the form of concentric spherical shell lightcones, which are separated by $\sim 70$~Myr. The maximum radial comoving distance in the simulation is $\simeq 3$~Gpc, which corresponds to a redshift of $z \simeq 1.4$. 
    
    This simulation has a volume that is $216$ times the volume of the Millenium simulation \citep[][]{millenium_simulation} and $24000$ times the Illustris simulation \citep[][]{illustris_1}. The MICE Onion Universe simulation also has more dark matter particles than the Illustris simulation, giving a better mass resolution in the simulation. These factors make the MICE Onion simulation best suited for large scale statistical analyses based on the dark matter distribution. The drawback of this simulation is the limited spatial resolution offered in the output dark matter maps with objects smaller than galaxy-sized halos being unresolved in this simulation. However, this is not a significant issue for our analysis as we are interested in the intergalactic DM contribution and not the DM contribution from individual galaxies or their halos. We note that any intervening halos along the line of sight will contribute to the DM and can do so significantly \citep{Prochaska2019}.
    
    These data were available in the form of HEALPIX maps \citep[][]{healpix}, with a resolution defined by $N_{\rm side} = 4096$, providing an angular resolution of $0.85'$ on the sky. The data set contains the dark matter comoving number density per pixel, in units of $(\mathrm{Mpc}/h)^{-3}$. There are a total of 265 such maps corresponding to redshifts ranging from $0 \lesssim z \lesssim 1.4$.
    
    \subsection{Free electron density and dispersion measure}
        
        To calculate the contribution to the DM from the IGM, we convert the co-moving dark matter number density, $n_{\rm dark}(z)$, in each dark matter map (labeled with subscript $i$) at a sky location given by coordinates $\phi$ and $\theta$, to a comoving free electron density, $n_{\rm e}(z)$, 
        \begin{eqnarray}
            \displaystyle n_{\rm e}(z_i|\phi,\theta) & = & \rho_{\rm bary}(z_i|\phi,\theta)  f_e \nonumber \\
            & = & \frac{n_{\rm dark}(z_i|\phi,\theta) \Upsilon f_e}{\Xi}.
        \label{free_e_density_eq}
        \end{eqnarray}
        where $\rho_{\rm bary}$ is the baryon matter density, $f_e$ is the electron ionization fraction, $\Upsilon$ is the ``mass factor" to convert the dark matter particle mass to the equivalent baryonic mass and $\Xi$ is the scaling factor between the dark and baryonic matter density.
        
        To do this, we first convert the dark matter number density, $n_{\rm dark}(z)$, to the corresponding baryonic matter number density, $n_{\rm bary}(z)$, by using the scaling relation between dark matter and baryonic matter energy density \citep{wmap},
        \begin{equation}
            \displaystyle \Xi \equiv \frac{\Omega_{\rm dark}}{\Omega_{\rm bary}} \approx \frac{26.9}{4.6} \approx 5.8,
            \label{dm_bary_rel}
        \end{equation}
        where $\Omega_{\rm dark}$ and $\Omega_{\rm bary}$ are the dark matter and baryonic matter energy density respectively. Since one dark matter particle of the simulation does not correspond to one baryonic matter particle, we need to add an additional ``mass factor", $\Upsilon$, to get the baryon matter density, $\rho_{\rm bary}(z)$. Assuming a universe containing only hydrogen and helium, the mass factor is
        \begin{equation}
            \displaystyle \Upsilon = \frac{M_{\rm dark}}{(M_{\rm H} + M_{\rm He})}, 
        \end{equation}
        where $M_{\rm H}$ and $M_{\rm He}$ are the mass of a hydrogen and helium atom respectively.
        
        Next, we convert the baryonic matter number density to a corresponding free electron density. This involves multiplication of the baryonic matter density by an electron ionization fraction, $f_e$. Using the formulation presented in \citet[][]{zheng_dm_igm}, we define the electron ionization fraction as,
        \begin{equation}
            \displaystyle f_e = \left[ (1 - Y) f_{\rm H II} + \frac{1}{4} Y (f_{\rm HeII} + 2 f_{\rm HeIII}) \right],
            \label{f_e}
        \end{equation}
        where $Y = 0.25$ is the mass fraction of helium, $f_{\rm HII}$ is the ionization fraction of hydrogen, $f_{\rm HeII}$ and $f_{\rm HeIII}$ are the ionization fractions of single and double ionized helium. In this work, we assume that the hydrogen and helium in the IGM are fully ionized, i.e. $f_{\rm HII} = f_{\rm HeIII} = 1$ and none of the helium is partially ionized, i.e. $f_{\rm HeII} = 0$. Given our redshift range, we do not expect significant variations in these parameters.
        
        Putting it all together, we obtain Eq.~\ref{free_e_density_eq} as
        \begin{eqnarray*}
            \displaystyle n_{\rm e}(z_i|\phi,\theta) & = & \rho_{\rm bary}(z_i|\phi,\theta)  f_e \nonumber \\
            & = & \frac{n_{\rm dark}(z_i|\phi,\theta)  \Upsilon  f_e}{\Xi}.
        \end{eqnarray*}
        
\begin{figure*}
            \centering
            \includegraphics[width=\textwidth]{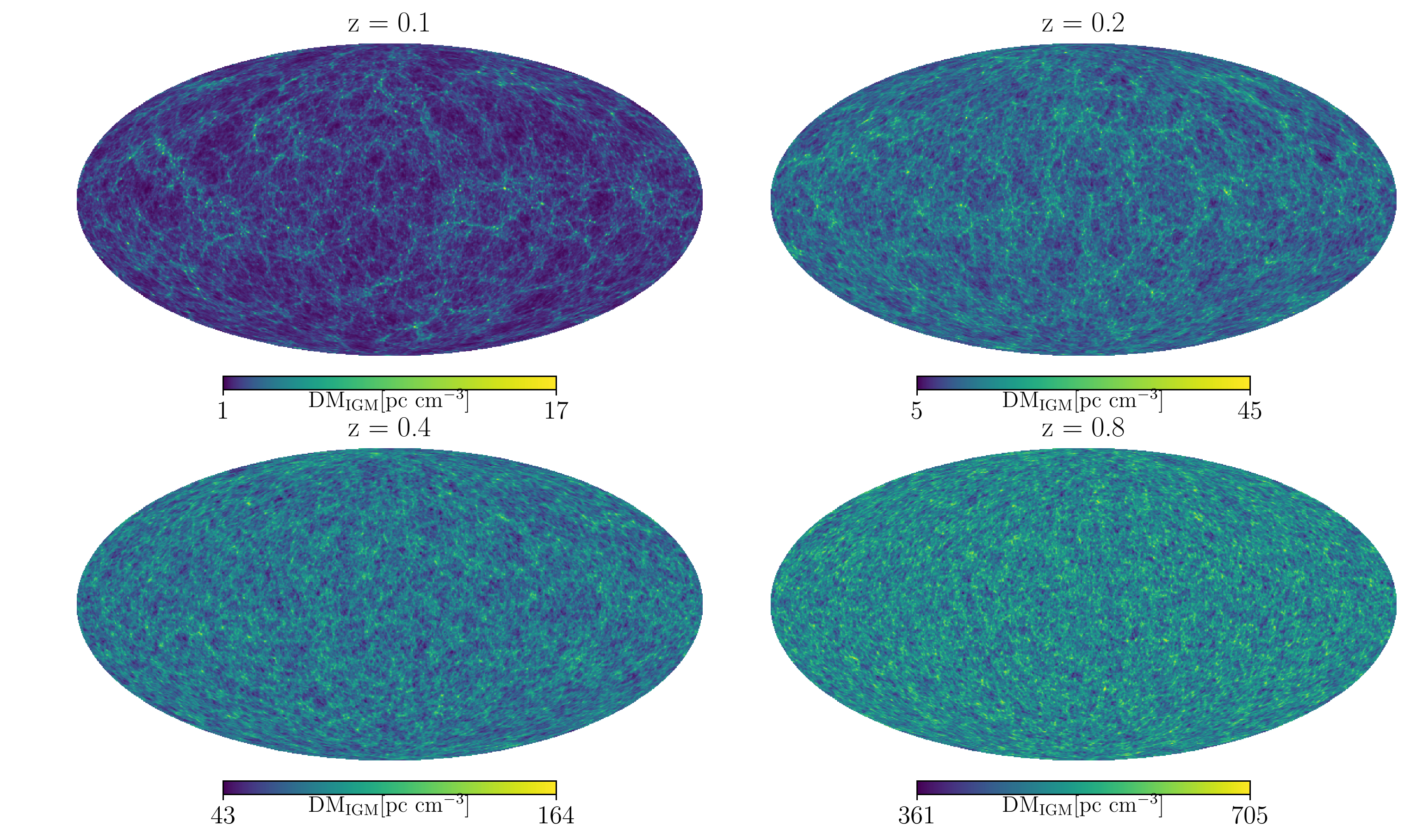}
            \caption{The intergalactic DM out to a redshift of $z = 0.1, 0.2, 0.4, 0.8$, in the top-left, top-right, bottom-left and bottom-right panels respectively. These maps are produced by smoothing the original maps with a symmetrical Gaussian beam for better visualization. Notice that the color-bar represents different ranges of the intergalactic DM in each map. We can see that at low redshifts, most of the intergalactic DM contribution is concentrated in filamentary structures on the sky, with the voids on the sky contributing small amounts of DM, which increases as we move to higher redshifts. An animated GIF image is made available with this paper which shows the variation in the $\DM_{\IGM}$ on the sky with respect to redshift.}
            \label{dm_map}
        \end{figure*}

        Finally, since each map has a finite width in redshift, $\Delta z$, we can directly convert the free electron density, $n_{\rm e}$, to the DM in each map,
        \begin{equation}
            \displaystyle  \DM_{\rm map}(z_i|\phi,\theta) = \frac{n_{\rm e}(z_i|\phi,\theta)}{1 + z_i}  \Delta l (z_i),
            \label{dm_eq}
        \end{equation}
        where the factor of $(1 + z_i)$ is introduced to account for the redshift in frequencies between the observer and source rest frame \citep{deng_frb}, and $\Delta l (z_i)$ is the comoving width of each map,
        \begin{equation}
            \displaystyle \Delta l = c \left| \frac{dt}{dz} \right| \Delta z
            \label{delta_l_eq}
        \end{equation}
        and 
        \begin{align}
        \begin{split}
            \displaystyle \left| \frac{dt}{dz} \right| = \frac{1}{(1 + z) H(z)} \\
            = \frac{c \ \Delta z}{H_0 (1 + z) (\Omega_m (1 + z)^3 + \Omega_{\Lambda})}
        \end{split}
        \end{align}
        We compute this comoving width for each map using  Astropy's\footnote{\url{http://www.astropy.org}} \citep[][]{astropy} Cosmology module. Thus, with the above recipe, we have the DM contribution from the intergalactic medium, $\DM_{\rm map}(z_i)$, at different redshifts. A few examples of these maps are shown in Fig.~\ref{dm_map}. Finally, to compute the integrated DM, $\DM_{\IGM}$, along each line of sight on the sky, we sum the DM contribution of each map out to the desired redshift,
        \begin{equation}
            \displaystyle \DM_{\IGM}(z|\phi,\theta) = \sum_{i = 1}^{z_i = z} \DM_{\rm map}(z_i|\phi,\theta)
            \label{final_dm_eq}
        \end{equation}
    
\section{Dispersion Measure Statistics} \label{stats_from_maps}

    Once we calculate the DM along every line of sight $(\phi,\theta)$ as given by Eq.~\ref{final_dm_eq}, we can construct probability distributions for the DM at a given redshift. We will look at two methods of constructing these probability distributions. The first is to create histograms of $\DM_{\IGM}(z)$ for all values of $(\phi,\theta)$. This is equivalent to assuming that the probability of an FRB being on any line of sight is the same. Next, we weight the histograms by assuming that the FRB distribution follows the matter distribution.
    
    \subsection{Uniform Weighting}
        
\begin{figure*}
            \centering
            \includegraphics[width = \textwidth]{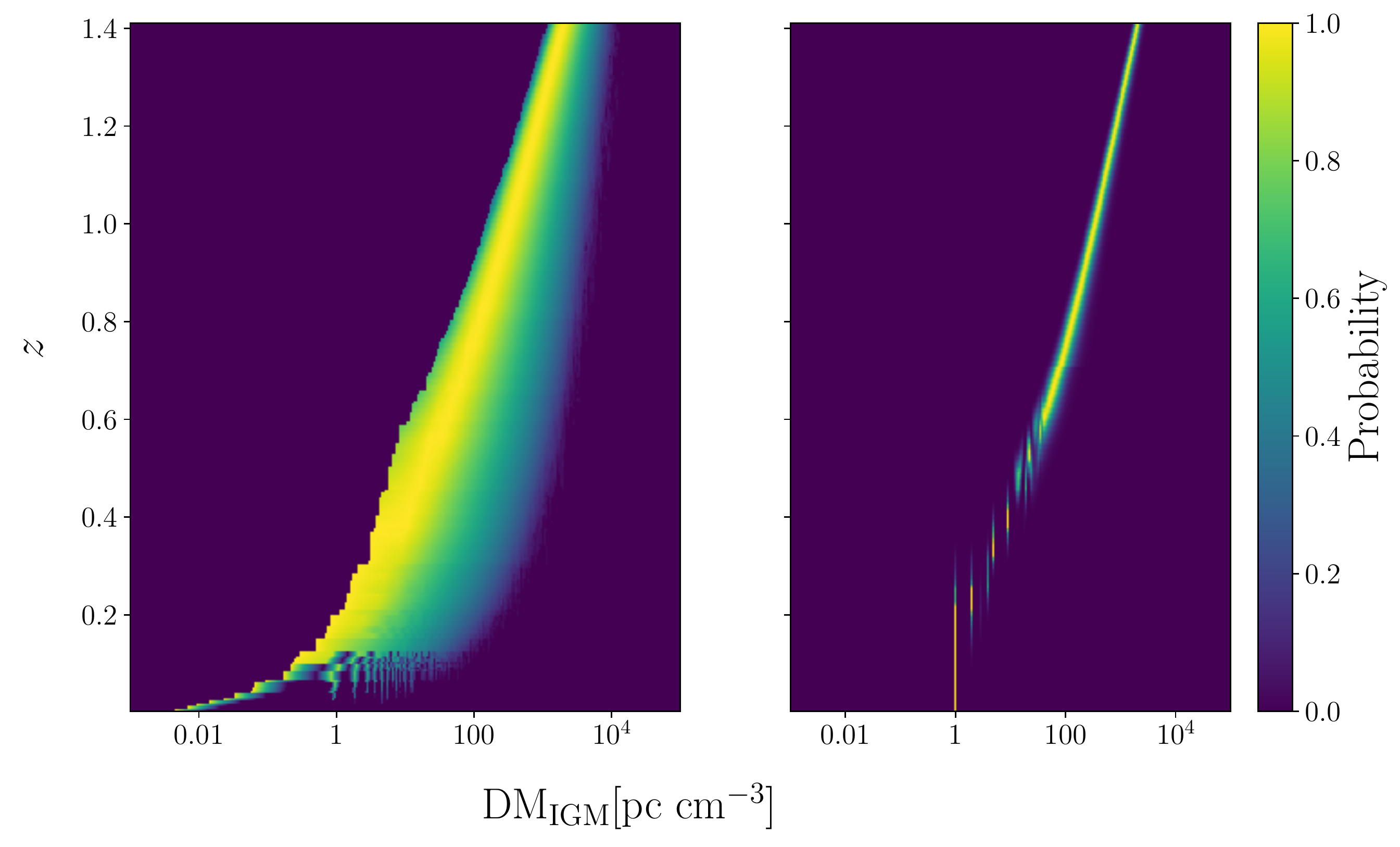}
            \caption{The intergalactic DM, $\DM_{\IGM}$, PDFs are plotted here as a function of redshift. The left panel represents the PDFs obtained using the uniform weighting scheme, while the right hand panel represents the PDFs obtained using the matter distribution weighting scheme. As described in Sec.~\ref{stats_from_maps}, the patchiness of the electron density causes the PDF of $\DM_{\IGM}$ to be multimodal at redshifts $z \leq 0.1$, which results in most of the sky having very low $\DM_{\IGM}$. As we move back in redshift, the distribution of $\DM_{\IGM}$ becomes more uniform across the sky, assuming a skewed distribution with long trailing edges visible in the uniform weighted PDFs.}
            \label{dm_dists}
        \end{figure*}

        The intergalactic DM probability distributions created by assuming an FRB at the end of every line of sight are shown in the left hand panel of Figure~\ref{dm_dists}. These PDFs are skewed towards higher values of DM due to the overdensities in the IGM contributing significantly higher DM than other regions of the IGM. At redshifts $z \lesssim 0.1$, the IGM is highly structured, which results in the majority of the sky making a small DM contribution. However, the structures in the IGM contribute significantly larger DMs, even though there are very few regions on the sky that contain these structures. This can be seen in the left-hand panel of Figure~\ref{dm_dists}, where the distributions for $z \lesssim 0.1$ are heavily concentrated around low values of DM, with the contribution from the IGM structures visible as the outliers at significantly high DM values. 
        
\begin{figure*}
            \centering
            \includegraphics[width = \textwidth]{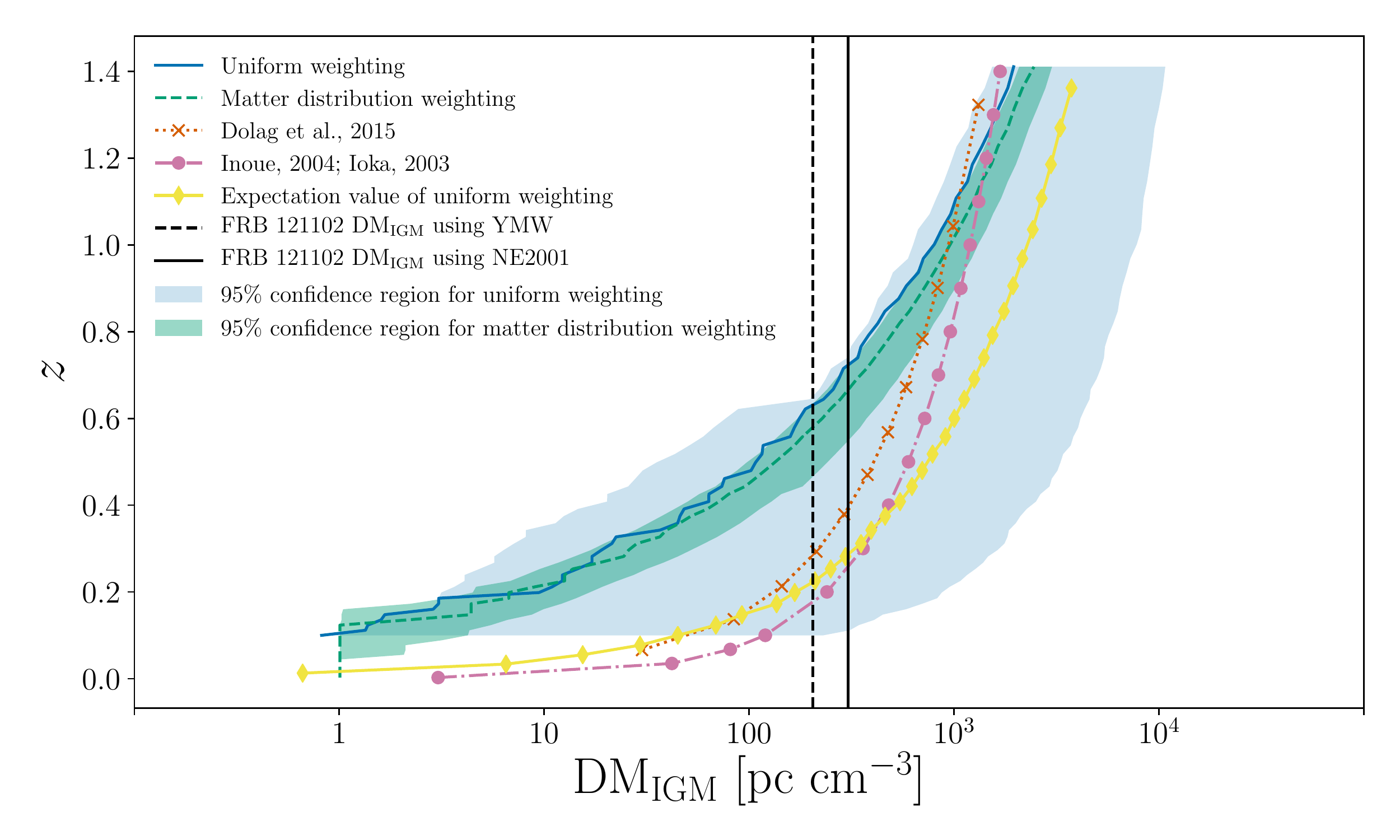}
            \caption{The redshift is plotted here against the most probable intergalactic DM. The most probable contribution is the peak of the distributions shown in Fig.~\ref{dm_dists}, with the distributions obtained using the uniform weighting scheme shown as the solid blue line and those obtained using the matter distribution weighting scheme shown as the dashed green lines. The shaded region represents the 95\% confidence interval, which is measured by excluding 2.5\% of the area at the head and tail of the distributions shown in Fig.~\ref{dm_dists}. If the peak of the distribution happens to coincide with the lowest bin, the lower limit on the confidence interval is set to be equal to the value of the peak. We also plot as the yellow solid line the expectation value of the intergalactic DM obtained from the uniform weighting scheme. For comparison, we also plot the intergalactic DM estimates made by \citet[][]{inoue_dm} and \citet{dolag_dm_igm} as pink dash-dot line and brown dashed lines respectively. The vertical dashed lines represent the $\DM_{\IGM}$ contribution for FRB 121102, assuming $\DM_{\host} = 0$ (see Eq.~\ref{dm_igm_eq}), when using the YMW model for the Galactic free electron density while the solid line represents the same but using the NE2001 model. As we can see, the two different models produce different redshift upper limits, but these are consistent within the large range of predicted $\DM_{\IGM}$ values. This also illustrates that other models depend more significantly on the choice of the Galactic free electron density because they have a much steeper redshift to $\DM_{\IGM}$ mapping.}
            \label{dm_v_z}
        \end{figure*}

        The most probable (i.e., maximum likelihood) contributions of the IGM to the DM with uniform weighting as a function of redshift are shown in Fig.~\ref{dm_v_z}. The solid blue line represents the peaks of the DM probability distributions shown in Fig.~\ref{dm_dists}, while the blue shaded region represents the 95\% confidence region. The expectation value of the distribution at each redshift is also shown as the yellow solid line in the same figure. As we can see, the most probable contribution of the IGM to the DM tends to be very low up to significantly high redshifts. The most probable $\DM_{\IGM}$ does not exceed 100~\pcc\ until a redshift of $z \sim 0.48$, while it does not exceed 1000~\pcc\ until a redshift of $z \sim 1.03$. However, we note that there are significant deviations from these most probable DMs depending on the line of sight along which the FRB might be located, as is evident from the distributions plotted in Fig.~\ref{dm_dists}. The expectation value of the distributions at each redshift is significantly higher than the most probable values due to the skewed distribution of the IGM DM distribution. We will discuss the significance of this in Sec.~\ref{comparison}.
        
    \subsection{Weighting by the Matter Distribution}

        We repeat our analysis by assuming that the probability of an FRB being at the end of a given line of sight is proportional to the matter density at the final redshift map. That is, rather than add one count to the histogram for each value of $\DM_\IGM(z|\phi,\theta)$, we add a weight equal to the number of dark matter particles, which is proportional to the dark matter particle number density, $n_{\rm dark}(z_i|\phi,\theta)$. This method assumes that FRBs are more likely to be coincident with higher matter concentrations (e.g., galaxy clusters) and less likely in lower matter concentrations (e.g., voids).
        
        The intergalactic DM probability distributions obtained in this fashion are shown in the right hand panel of Fig~\ref{dm_dists}. We can see that following this approach leads to a much tighter distribution of the possible intergalactic DMs than for the case of uniform weighting. Similarly, we plot the most probable contribution of the IGM under this weighting scheme as the dashed green line in Fig.~\ref{dm_v_z}, with the shaded green region representing the 95\% confidence region.
        
        As we can see in Fig.~\ref{dm_v_z}, the most probable contribution of the intergalactic DM under this weighting scheme closely tracks the contribution obtained under the uniform weighting scheme, but with much tighter constraints on the most probable DM values. The tighter constraints on the upper bounds of the intergalactic DM suggest that this weighting scheme tends to largely ignore the contribution from the structure in the IGM, which is represented by the long tails of the probability distributions obtained in the uniform weighting scheme. This is to be expected, since the number of pixels, and hence the total number of particles, belonging to the structure at any given redshift are small in number compared to the majority of the pixels which do not belong to this structure. 
        Thus, the relative paucity of these pixels results in them having a smaller overall weight as compared to the pixels which do not belong to the structures. This results in the final PDF de-emphasizing the contribution from the structures, or tails, of the PDF obtained using the uniform weighting scheme.
        
        In the rest of the paper, we report values from both of the weighting schemes wherever it is appropriate.
        
    \subsection{Comparison with Other $\DM_{\IGM}$ Predictions} \label{comparison}
        
        There have already been a number of estimates of $\DM_{\IGM}$ relying on either an analytical approach or simulations of the IGM. In this section, we compare the results from both of these approaches with those that we obtain in this work.
        
        The analytical estimates in general rely upon expressions which track the evolution of the free electron density as a function of redshift. An analytical estimate of the $\DM_{\IGM}$ was made by \citet[][]{ioka_dm} and \citet[][]{inoue_dm}, whose model predicted a $\DM_{\IGM}(z=1) \sim 1200$~\pcc~ (Eq.~\ref{inoue_reln}). A similar estimate made by \citet[][]{zhang_dm_igm} predicts $\DM_{\IGM} (z = 1) \sim 855 \pm 345$~\pcc. In comparison, based on our analysis with uniform weighting, we derive a $\DM_{\IGM} (z = 1) \sim 800^{+7000}_{-170}$~\pcc, and with weighting by matter distribution, we derive a $\DM_{\IGM} (z = 1) \sim 960^{+350}_{-160}$~\pcc, where the errors represent the 95\% confidence interval. 
        
        As we can see in Fig.~\ref{dm_v_z}, the $\DM_{\IGM}$ estimate made in our work using the uniform weighting scheme is consistent with the estimates made in these other works within our confidence intervals. However, the other works like \citet[][]{ioka_dm} and \citet[][]{inoue_dm} predict higher values of $\DM_{\IGM}$ at low redshifts relative to our most probable values, i.e. the matter distribution weighting scheme. The discrepancy can be partly explained by the fact that these studies are based on calculating the mean or expectation values of the $\DM_{\IGM}$ at each redshift from analytical expressions. As we show in this work, especially in Fig.~\ref{dm_dists}, if the underlying probability distribution of the $\DM_{\IGM}$ is skewed and the analytical approach does not account for this, the most probable value of the intergalactic DM at any redshift will be systematically lower than the expectation value of the distribution. If we calculate the expectation value of $\rm log(\DM_{\IGM})$ (treating $\DM_{\IGM}$ as a log-normal random variable)  at each redshift in the uniform weighting scheme, then we obtain values at low redshift that are more consistent with those obtained by other works as shown in Fig.~\ref{dm_v_z}. 
        In addition, the studies based on analytical expressions are representative of the average behavior of the IGM and do not take into account the spatial variation of the free electron density, and thus the DM, at a given redshift. As a result, these studies might underestimate the errors on the $\DM_{\IGM}$ values.
        
        The other approach for calculating $\DM_{\IGM}$ is using large cosmological simulations of dark matter and/or baryonic matter. \citet[][]{mcquinn_dm_igm} used this approach to examine the spatial variation and found that from redshifts 0.5 to 1, the standard deviation about the mean was between 100 and 400~\pcc~(which must be multiplied by $\sim$2 to arrive at the inner 95\% confidence interval assuming Gaussian statistics) both from analytical expressions and also via a 40 Mpc/$h$, $2 \times 512^3$ dark matter particle cosmological simulation \citep[][]{FaucherGiguere_dark_matter_sim} to arrive at the baryon distribution for halos along lines of sight.
        Another estimate of the intergalactic DM contribution was made by \citet[][]{dolag_dm_igm}, where they used \textit{Magneticum Pathfinder}\footnote{\url{http://www.magneticum.org/}} \citep[][]{dolag_dm_igm} simulation, with a box size of 896 Mpc/$h$, which is approximately a factor of 3.4 times smaller than the MICE Onion simulation used here, to simulate the cosmic web, which includes both dark matter and gas matter particles. To compare our results with theirs, we calculate the peaks of their $\DM_{\IGM}$ (they use the term $\DM_{\rm cosmo}$ to represent the same) distributions in the left-hand panel of Fig.~4 using Eqs.~6--10 \citep[see][]{dolag_dm_igm} and plot them alongside our results in Fig.~\ref{dm_v_z}. A similar cosmological-simulation-based estimate of the intergalactic DM contribution was also made by \citet{illustris_dm_igm} using the Illustris simulation \citep{illustris_1}, where they directly integrated the free electron density over the box size of 75 Mpc/$h$, over a factor of 100 smaller than the MICE Onion simulation used here, though containing baryonic physics. In that study, they predict a $\DM_{\IGM}(z = 1) = 905 \pm 115$~\pcc~(error represents one standard deviation). The $\DM_{\IGM}$ estimates from both of these analyses are consistent with the ones we obtain with the uniform weighting scheme within the confidence intervals.
        
        However, as with the analytical studies, at low redshift, the $\DM_{\IGM}$ estimates from \citet[][]{dolag_dm_igm} are larger than the most probable values that we calculate. This can be explained by the fact that the simulation used by \citet[][]{dolag_dm_igm} has a volume 40 times smaller and spatial resolution approximately an order of magnitude smaller than our underlying MICE simulation. In addition to this, the mass of the gas matter particles in their simulation is a factor of 10 smaller than in the MICE simulation. Put together, this implies that \citet[][]{dolag_dm_igm} are able to resolve dense structures like galaxy groups and halos (which are unresolved in the MICE simulation). Consequently, this will increase the average particle density in their simulation, which will in turn shift the calculated $\DM_{\IGM}$ towards higher values.
        
        In our work with the MICE simulation, we are probing structures on much larger scales which the \citet[][]{dolag_dm_igm} and \citet[][]{illustris_dm_igm} simulations are insensitive to, and which tend to be rarer and thus have lower spatial densities. As a result, the inferred $\DM_{\IGM}$ distribution will peak at lower values than that for higher resolution surveys like the one from \citet[][]{dolag_dm_igm} and the Illustris simulation \citep[][]{illustris_dm_igm}. As noted in \citet[][]{dolag_dm_igm}, another consequence of probing much larger scale structures is that it allows us to accurately model the tail of the $\DM_{\IGM}$ distributions better than \citet[][]{dolag_dm_igm}, thereby providing a more complete representation of the range of $\DM_{\IGM}$ values at a given redshift.
        
        Finally, as we move towards higher redshifts the structure that is seen at low redshifts begins to homogenize. In simulations such as the ones used by \citet{dolag_dm_igm} and \citet[][]{illustris_dm_igm}, this implies that the fraction of baryons in the IGM increases towards higher redshifts, i.e. the particle density attributed to galaxies and their halos decreases, while that in the IGM increases. As a result, we see the $\DM_{\IGM}$ values measured in \citet[][]{dolag_dm_igm} converge towards values that are similar to the most probable values that we obtain in this work.
        
        The above discussion highlights the difference due to using a single number over the entire distribution of the $\DM_{\IGM}$ in making predictions based on the $\DM_{\IGM}$. For example, estimates of the redshift of the FRB made using the expectation values would tend to be underestimates as compared to using the most probable values. Since redshift predictions usually involve the assumption of $\DM_{\host} = 0$, the estimates based on these studies are interpreted as upper limits. Not accounting for the entire possible range of redshifts might lead to follow-up searches not exploring the entire redshift space when searching for the FRB host galaxy in archival data. Another consequence of the use of only the expectation values over the entire distribution of $\DM_{\IGM}$ would result in underestimating the host galaxy contribution to the total DM. This might hinder the efforts to understand the local environment of the FRB (or its host galaxy) based on the inferred DM values (see Sec.~\ref{askap_loczn} for an example).
        
        We performed a maximum-likelihood fit for the most probable $\DM_\IGM$ as shown in Fig.~\ref{dm_v_z} with a parabolic curve in log-log space along with an additional parameter to determine the uncertainty in the relation. We find the functional form to be
        \be 
        \log_{10}\DM = 0.48(\log_{10}z)^2 + 3.34(\log_{10}z) + 2.98
        \label{eq:functionalform}
        \ee
        where DM is in units of \pcc\ and the error in the relation is $\sigma_{\log_{10}\DM} = 0.09$. 
        
        We used the uniform-weighting curve because it avoids the flattening at $z \lesssim 0.1$, though qualitatively the curves match quite well at all higher redshifts. This equation can be used to extrapolate to higher redshifts though of course care should be taken in doing so.
        
    \subsection{IGM DM contribution under 100 Mpc}
        
        The galaxy clusters closest to the Milky Way, such as the Virgo, Fornax and Hydra clusters \citep{local_clusters}, lie within 100~Mpc \citep[$z = 0.023$,][]{wright_cosmo_calc} of the Milky Way \citep{laniakea_paper}. The intergalactic DM contribution at these low redshifts is small relative to the contribution from the Milky Way and the host galaxy. For example, the Milky Way ISM contribution (not including the halo contribution) in the direction of the Virgo cluster, which is at a distance of approximately 20~Mpc \citep[$z \approx 0.004$,][]{wright_cosmo_calc}, is $\DM_{\ISM} \approx 30$~\pcc\ \citep{ne2001}, while the intergalactic DM contribution, based on our work, is $\DM_{\IGM} < 4$~\pcc.
        
        Because of these low DMs involved, care must be taken about the range of DMs over which FRBs are searched for in these nearby clusters.
        
\section{Redshift limits on fast radio bursts} \label{redshift_secn}
    
    One application of this analysis is in deriving redshift limits for the observed FRBs. Using Eq.~\ref{eq:DMmodel}, we can calculate the empirical $\DM_{\IGM}$,
    \begin{equation}
        \displaystyle \DM_{\IGM}(z) = \DM_{\rm obs} - \DM_{\ISM} - \DM_{\CGM} - \frac{\DM_{\host}}{1 + z}
        \label{dm_igm_eq}
    \end{equation}
    Given the lack of modeling of the host galaxy DM, we assume $\DM_{\host} = 0$~\pcc~for now. This implies that the redshift limits we derive will be upper limits.
    
    We can compute the joint probability distribution function (PDF), $f(\DM_{\IGM}, z)$, by normalizing the two dimensional histogram shown in Fig.~\ref{dm_v_z}. Using this joint PDF and an intergalactic DM, $\DM_{\IGM}$, deduced from Eq.~\ref{dm_igm_eq}, we can estimate the probability, $f(z|\DM_{\IGM})$, that the FRB lies at a redshift, $z$, by
    \begin{multline}
        \displaystyle f(\DM_{\IGM}, z) = f(z|\DM_{\IGM}) \times f(\DM_{\IGM}) \\
        \implies f(z|\DM_{\IGM}) = \frac{f(\DM_{\IGM}, z)}{f(\DM_{\IGM})}, \\
        \label{z_prob}
    \end{multline}
    where $f(\DM_{\IGM})$ accounts for the error on the intergalactic DM deduced from Eq.~\ref{dm_igm_eq}, which includes the error on the measured total DM and the Milky Way contribution to the DM and therefore includes the observational uncertainties along with the uncertainties from the Milky Way electron density model and the CGM. 
    
\begin{figure*}[p!]
        \centering
        \includegraphics[width = \textwidth, height=\textheight, keepaspectratio]{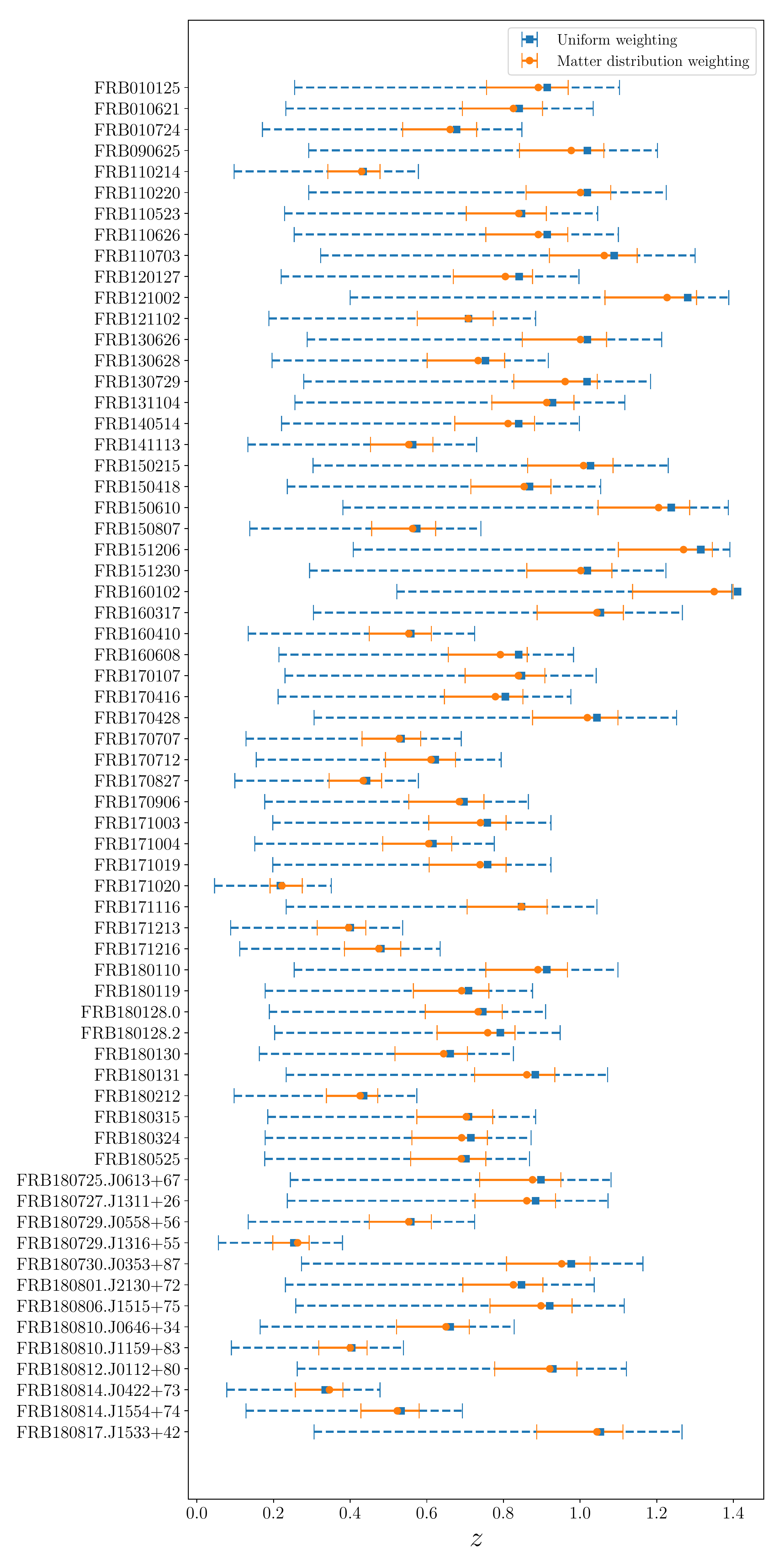}
        \caption{The 95\% redshift upper limits on all currently known FRBs using our model of the intergalactic DM, as described in Sec.~\ref{redshift_secn}, assuming $\DM_{\host} = 0$. The redshift limits based on the uniform weighting scheme are shown in blue, while those based on the matter distribution weighting scheme are shown in orange. These predicted redshifts imply that all FRBs detected so far are likely to be extragalactic sources. While we can place constraining upper limits on the redshifts of other FRBs, we note that the DM measured for FRB 160102 is high enough that we can only place a lower limit on the possible redshift of this source.}
        \label{redshifts}
    \end{figure*}

    It is well known that both the NE2001 and YMW models have difficulties modeling the free electron density in certain directions \citep[see, for example,][]{wrong_dmdist_1, wrong_dmdist_2, wrong_dmdist_3}. In the following, we use the NE2001 model assuming a conservative 50\% error \citep{ne2001} on the Milky Way's DM contribution predicted using this model. We repeated the analysis with the YMW model, but did not find any significant differences in the results because our mapping of the redshift to $\DM_{\IGM}$ is much shallower than that for other works, as explained in Fig.~\ref{dm_v_z}. Thus we do not report results from using the YMW model, but provide the reader the option to choose between the NE2001 and YMW model in the software \citep{our_code} provided with this paper\footnote{We provide Python scripts to generate the probable redshift of an FRB given its total DM and position on the sky in the GitHub repository associated with this work: \url{https://github.com/NihanPol/DM_IGM}}.
    For the Halo contribution, we use the 50-80~\pcc\ range from \cite{prochaska_dm_cgm} and then assume that the Halo contribution is $65 \pm 15$~\pcc. For all three values, we assume that the probabilities are given by a Gaussian distribution with standard deviation equal to the appropriate errors. Then, using Eq.~\ref{z_prob}, we place upper limits on the redshifts of the FRBs using both the uniform weighting and matter distribution weighting, shown in Fig.~\ref{redshifts}. While we can place constraining upper limits on the redshifts of other FRBs, we note that the DM measured for FRB 160102 is high enough that we can only place a lower limit on the possible redshift of this source. However, a more reasonable host-galaxy DM contribution could be used to reduce the amount of DM attributed to the IGM, which would in turn allow us to place better constraints on the redshift upper limit.
    
    The redshift upper limits predicted using matter distribution weighting are much tighter than those predicted using uniform weighting. This is because the intergalactic DM distributions generated using the matter distribution weighting scheme are much narrower. Similarly, the larger errors on the redshifts predicted using uniform weighting, especially those towards lower redshifts, are a result of the long tails of the intergalactic DM PDFs shown in Fig.~\ref{dm_dists}. In our opinion, the redshifts predicted using the matter distribution weighting can be used as upper limits for any given FRB.
    
    We can use the redshift for FRB 121102 \citep[$z = 0.19$,][]{121102_localzn_2} to compare our predictions to the true redshift as well as those from other models. Using the uniform weighting scheme for the intergalactic DM distribution, we predict a redshift range of $z_{121102, {\rm uniform}} = 0.68_{-0.52}^{+0.17}$, while using the matter distribution weighting scheme, we predict a redshift limit of $z_{121102, {\rm matter}} = 0.67_{-0.13}^{+0.07}$. Both the weighting schemes predict a redshift higher than the true redshift of FRB 121102, which is to be expected since these predictions assume a host contribution of $\DM_{\host} = 0$~\pcc. As we show in Sec.~\ref{host_dm_secn}, the host galaxy contribution for FRB 121102 (and probably all FRBs) is significant. Thus, we can treat $z_{121102} \leq 0.74$ as an upper limit on the redshift of FRB 121102.
    
    We can then compare our redshift upper limits to those made in other works, especially those by \citet{ioka_dm, inoue_dm} who predict an upper limit of $z \leq 0.32$. Accounting for the variance in mapping the DM to redshift \citep{mcquinn_dm_igm} increases this upper limit to $z \leq 0.42$. These upper limit predictions are smaller than those that we make in this work and might result in searches for the host galaxy in archival data not exploring the entire possible redshift space, as described in Sec.~\ref{comparison}.
    
    
    We can make similar comparisons for the two other FRBs that have been localized in redshift. For FRB 180924 \citep[z = 0.32,][]{askap_loczn}, we predict an upper limit of $z_{180924, {\rm matter}} \lesssim 0.62^{+0.06}_{-0.12}$, while using the \citet{ioka_dm}, \citet{inoue_dm} method results in a redshift upper limit of $\lesssim$0.25 and accounting for the variance suggested by \citet{mcquinn_dm_igm} increases that to $\lesssim$0.35. Similarly for FRB 190523 \citep[$z = 0.66$,][]{dsa10_loczn}, we predict an upper limit of $z_{190523, {\rm matter}} \lesssim 0.87^{+0.06}_{-0.15}$, and using the \citet{ioka_dm}, \citet{inoue_dm} method results in a redshift upper limit of $\lesssim$0.56, which increases to $\lesssim$0.66 when accounting for the \citet{mcquinn_dm_igm} variance. Thus, the upper limits predicted using \citet{ioka_dm}, \citet{inoue_dm} method are in tension with the measured redshift for FRBs 180924 and 190523 and require the \citet{mcquinn_dm_igm} variance to produce upper limits consistent with the measure redshifts. Our upper limits are not susceptible to this problem and provide better estimates of the resdhift upper limit.
    
    Finally, our limits can be used to eliminate potential host galaxies that fall above our predicted upper limit on the redshift and can reduce the time required to search for a host galaxy. This will be useful for large field-of-view surveys which do not have arcsecond localization capabilities and thus have lots of potential host galaxies in their field-of-view, for example, the Canadian HI Intensity Mapping Experiment (CHIME) FRB experiment \citep[][]{chime_frb}. 

\section{Estimating the host galaxy dispersion measure} \label{host_dm_secn}
    
    Another application of this work is to allow direct estimates of the host galaxy DM contribution, $\DM_{\host}$, if the redshift of the FRB, and thus the host galaxy, is known. Again using Eq.~\ref{eq:DMmodel}, we can directly calculate the PDF for the DM of the host galaxy,
    \begin{multline}
        \displaystyle \DM_{\host} = (\DM_{\rm obs} - \DM_{\ISM} \\
        - \DM_{\CGM} - \DM_{\IGM}) \left(1 + z \right)
        \label{dm_host_eq}
    \end{multline}
    This $\DM_{\host}$ is a sum of the DM local to the source of the FRB and the DM contributed by the interstellar medium in the rest of the host galaxy along the line of sight. For non-repeating FRBs, the observed DM, $\DM_{\rm obs}$ will be a single measured value, while for repeating FRBs, $\DM_{\rm obs}$ will be a distribution depending on the variation of the total measured DM at different epochs. These short-timescale DM variations are likely to be local to the source of the FRB and will be reflected as a broadening in the $\DM_{\host}$ PDF calculated using Eq.~\ref{dm_host_eq}. We provide Python code to do this calculation in the same repository that hosts code to calculate the redshift limits for the FRBs \citep{our_code}\footnote{\url{https://github.com/NihanPol/DM_IGM}}.
    
    \begin{figure}
        \centering
        \includegraphics[width = \columnwidth]{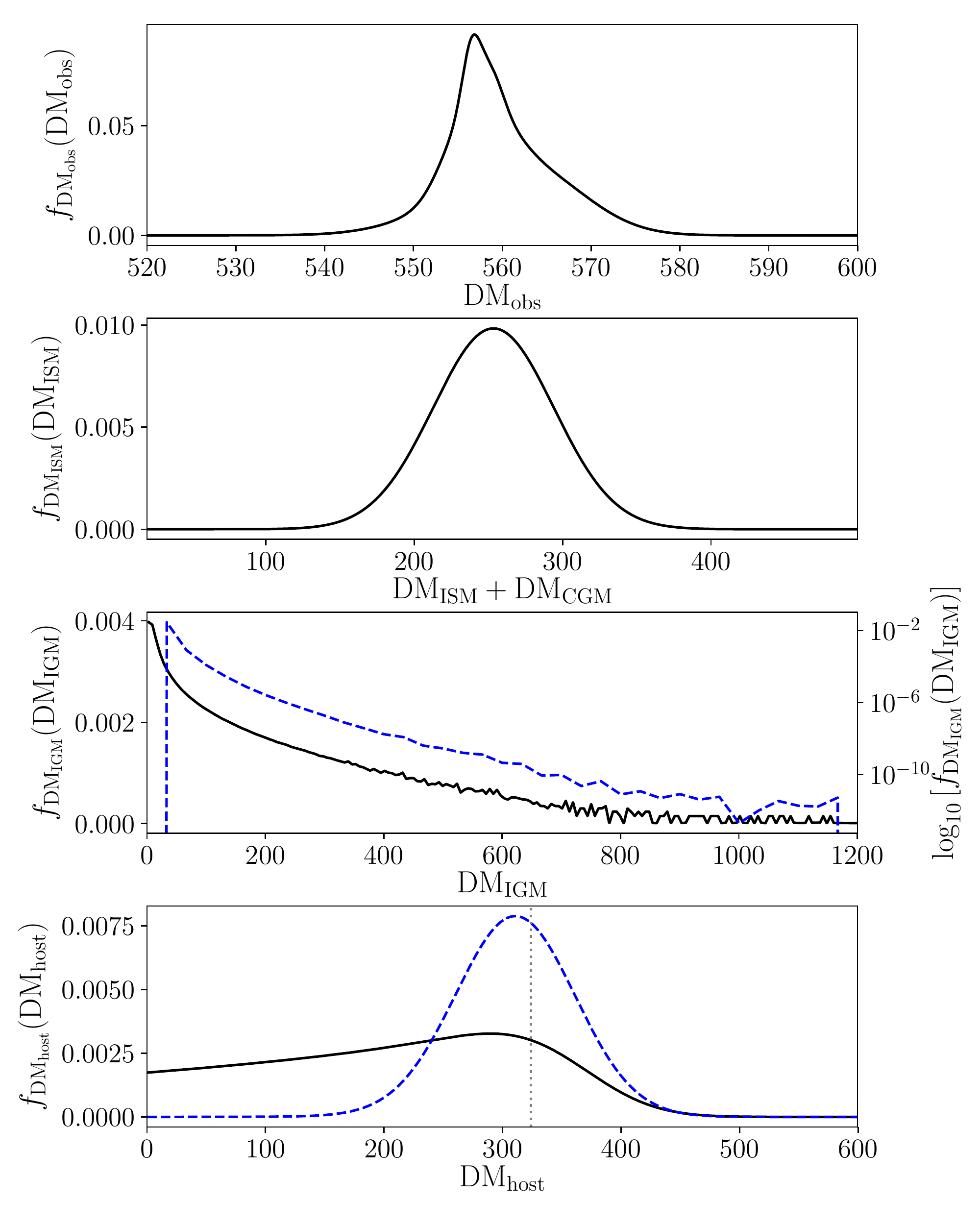}
        \caption{\textit{Top panel}: The distribution of the observed total DM for FRB 121102 generated using the DMs reported in \citet{121102_repeats} and \citet{121102_dm_vary_1}. \textit{Second panel}: The sum of the Milky Way \citep{ne2001} and circumgalactic medium \citep[halo,][]{prochaska_dm_cgm} contributions to the total DM for FRB 121102. We assume a 20\% error on the Milky Way contribution to the total DM, while we assume a circumgalactic DM contribution of $65 \pm 15$~\pcc. \textit{Third panel}: The intergalactic DM from the redshift slice at $z = 0.19$ \citep{121102_localzn_2} generated in this work, where the black now denotes the uniform weighting scheme and the dashed blue lines show the distributions using the matter distribution weighting. Note the wide range of scales involved for the matter distribution weighted distribution, shown on a logarithmic axis on the right. \textit{Bottom panel}: The host galaxy DM contribution to the total DM. This is obtained by subtracting the Milky way, circumgalactic, and IGM DM distributions from the total DM distribution. As before, the dashed blue line represents the PDF of the matter distribution weighted distribution, though now on a linear scale. The dotted gray line shows the Balmer-line-derived DM of 324~\pcc\ \citep{121102_localzn_2}. Note the varying scales on the horizontal axes and that the units are all in the standard pc~cm$^{-3}$.}
        \label{host_dm}
    \end{figure}
    
    \subsection{FRB 121102}
        
        The PDFs of the different contributions to the DM are shown in Fig.~\ref{host_dm} using the example of FRB 121102 \citep{121102_disc}. Multiple bursts ($\sim$100) have now been detected from FRB 121102 \citep[for example,][]{121102_dm_vary_1, 121102_dm_vary_2, 121102_ml_bursts}. All the bursts have been observed with an average DM, $\DM_{\rm obs} = 557$~\pcc, though different bursts have shown different DMs \citep[][]{121102_repeats, 121102_dm_vary_1}. Even accounting for the time-frequency structures of the bursts and correcting the DMs, \cite{Hessels+2018} find true DM variations over time as well. We combine the DMs reported for FRB 121102 in \citet[][]{121102_repeats} and \citet[][]{121102_dm_vary_1} to produce the conservative distribution of the observed DM given their wide spread and the use of determining the DM from the peak signal-to-noise ratio for the bursts; this distribution is shown in the top panel of Fig.~\ref{host_dm}.
        
        We compute the Milky Way contribution using the NE2001 model \citep{ne2001} which turns out to be 188~\pcc. We again assume a 20\% error associated with this Galactic DM contribution \citep{ne2001} and a circumgalactic DM contribution of $65 \pm 15$~\pcc\ to the Galactic DM to get the total DM contribution from the Milky Way, i.e. $\DM_{\ISM} + \DM_{\CGM}$, which is shown in the second panel in Fig.~\ref{host_dm}. Finally, using the fact that FRB 121102 lies at a redshift of $z = 0.19$, we choose the map at this redshift and use the PDF of the $\DM_{\IGM}$, both with uniform and matter distribution weighting, in this map as the intergalactic DM contribution. These PDFs of $\DM_{\IGM}$ are shown in the third panel in Fig.~\ref{host_dm}. Given these distributions, we can calculate the host galaxy contribution to the DM using Eq.~\ref{dm_host_eq}. The host galaxy DM, $\DM_{\host}$, so obtained is shown in the bottom panel of Fig.~\ref{host_dm}.
        
        As shown, we derive a broad range of possible $\DM_{\host}$, with the most probable contribution for the uniform weighting scheme being $\approx$280~\pcc, with an upper limit of 340~\pcc\ at the 95\% confidence level. For the matter distribution weighting, we find that $\DM_{\host} = 310 \pm 60$~\pcc. The latter is inconsistent with the estimates of \cite{Yang+2017}, where they estimated that host contribution was 210~\pcc. However, they assumed a Halo contribution of 30~\pcc\ and therefore comparing to our estimates using $\approx 65$~\pcc, their estimates are well below ours. As seen in Fig.~\ref{host_dm}, we find that both of our values are consistent with that calculated using the Balmer emission lines in the spectrum of the host galaxy. Converting the emission measure ($\EM = \int n_e^2(l)~dl$) to a DM by assuming strong variations in the electron density, \citet[][]{121102_localzn_2} approximate the host of FRB 121102 to have $\DM_{\rm Balmer}=324$~\pcc, though this value is highly uncertain due to the unknown nature of the host environment. 
        
        We can instead invert the analysis to constrain properties of the host galaxy medium. Following \citet{121102_localzn_2} \citep[see also][]{Cordes+1991,Cordes+2016}, we have that the DM value of the host galaxy with total path length $L$ should be 
        \ba
        \displaystyle \DM_\host & \approx & 948~\mathrm{pc~cm}^{-3}~\left(\frac{L}{6~\mathrm{kpc}}\right)^{1/2} \nonumber \\
        & & \times \left[\frac{f}{\zeta (1 + \epsilon^2) / 4} \right]^{1/2} \left(\frac{\EM_\host}{600~\mathrm{pc~cm}^{-6}}\right)^{1/2}\!\!\!\!
        \ea
        where $f$ is the filling factor of ionizing material in structures (e.g., clouds), $\zeta = \langle n_e^2\rangle/\langle n_e\rangle^2$ describes the electron-density variations between structures ($\zeta = 2$ for 100\% variations), and $\epsilon$ is the fractional density fluctuation (root-mean-square density divided by the mean density) within structures ($\epsilon = 1$ for fully modulated). The fiducial value for $\EM_{\host} \approx 600$~\pccc\ comes from the measurement in \citet[][]{121102_localzn_2}, as does the size of 4~kpc given the maximum diameter extent of the host galaxy they measure. Given our estimate of $\DM_\host = 310$~\pcc, we then have that 
        \be 
        \frac{f}{\zeta (1 + \epsilon^2) / 4} \approx 0.11.
        \ee
        Since by definition $f$ and $\epsilon$ cannot be greater than unity, and $\zeta$ must be greater than unity, then the filling factor along the line must be $\lesssim 4.4$\%, though it will likely be much less.
    
    \subsection{FRB 180924} \label{askap_loczn}
        
        FRB 180924 has only shown a single burst, but has been localized to a redshift of $z = 0.32$ and an early-type spiral galaxy with stellar mass of $2.2 \times 10^{10}$~M$_{\odot}$ \citep{askap_loczn}. Using the process described above for FRB 121102, we derive a host galaxy DM contribution for FRB 180924 to be $\DM_{\host, \rm uniform} = 210^{+40}_{-202}$~\pcc\ using the uniform weighting scheme and $\DM_{\host, \rm matter} = 220^{+36}_{-47}$~\pcc\ using the matter distribution weighting scheme.
        
        When trying to account for the DM budget (Eq.~\ref{eq:DMmodel}) for this FRB ($\DM_{\rm obs} = 361$~\pcc), \citet{askap_loczn} noticed that using the approach based on \citet[][]{ioka_dm}, \citet[][]{inoue_dm} or \citet[][]{prochaska_dm_cgm} for modeling the IGM DM resulted in the sum of the DM contribution from the Milky Way ($\DM_{\ISM} = 40$~\pcc), Milky Way halo ($\DM_{\CGM} = 60$~\pcc) and IGM \citep[$\DM_{\IGM} = 307$~\pcc,][]{prochaska_dm_cgm} exceeding the total measured DM by 46~\pcc. As we can see from Fig.~\ref{dm_v_z}, other IGM DM models predict a contribution that is about an order of magnitude higher than the most probable contribution from the IGM at this redshift based on our work. Since we account for the full IGM DM distribution at a given redshift in our model, our approach is not susceptible to these problems.
        
        To account for this over-budget DM, \citet[][]{askap_loczn} used the formalism developed by \citet[][]{mcquinn_dm_igm} to model the uncertainties in the IGM DM. Using this approach, they were able to satisfy the DM budget by deriving a host galaxy DM of $\DM_{\host} = 30-81$~\pcc, with 95\% upper limits of $77-133$~\pcc\ respectively. Using this approach results in a much lower host galaxy DM contribution for this FRB than that for FRB 121102. The host galaxy DM values derived using our model with uniform weighting scheme are consistent with the values derived by \citet{askap_loczn}, but are also consistent with the host galaxy DM for FRB 121102 which would allow for relatively similar local environments.
        
    \subsection{FRB 190523} \label{dsa10_loczn}
        
        Similar to FRB 180924, FRB 190523 has also shown a single burst and has been localized to a redshift of $z = 0.66$ and a galaxy with stellar mass of approximately $1.1 \times 10^{11}$~M$_{\odot}$ \citep{dsa10_loczn}. For this FRB, we derive a host galaxy DM contribution of $\DM_{\host, \rm uniform} = 420^{+115}_{-402}$~\pcc\ using the uniform weighting scheme and $\DM_{\host, \rm matter} = 401^{+67}_{-184}$~\pcc\ using the matter distribution weighting scheme.
        
        Just as for FRB 180924, using the \citet[][]{ioka_dm}, \citet{inoue_dm} or \citet[][]{shull_dm_igm} based approach implies a very small host galaxy DM contribution for this FRB \citep[$\DM_{\rm obs} = 760$~\pcc, $\DM_{\ISM} = 37$~\pcc, $\DM_{\CGM} = 50-80$~\pcc, $\DM_{\IGM} = 660$~\pcc,][]{dsa10_loczn}. Including an rms scatter of $200$~\pcc\ on the IGM DM contribution based on the work by \citet[][]{mcquinn_dm_igm} increases the upper limit on the host galaxy DM contribution to $\approx$210~\pcc. This estimate is consistent with the host galaxy DM calculated using our model with the uniform weighting scheme.
        
    \subsection{Host galaxy DM statistics}
        
\begin{figure}
            \centering
            \includegraphics[width = \columnwidth]{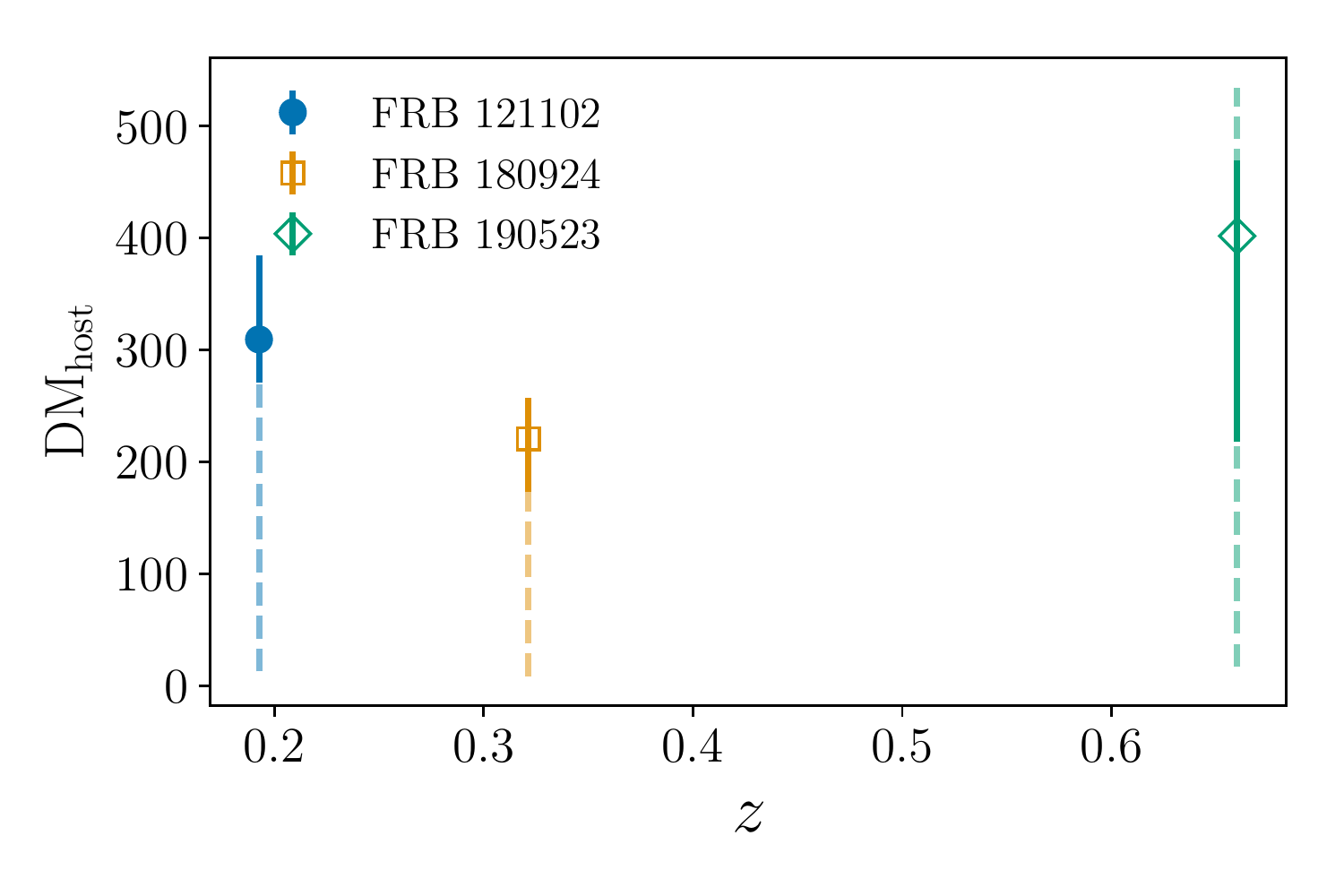}
            \caption{The host galaxy DM is plotted against the redshift of the host galaxy of the FRB. The solid lines represent the errors on the host galaxy DM obtained from the matter density weighted distribution while the dashed lines represent the errors obtained from the uniform weighting scheme. The solid marker represents the repeater FRB 121102, while the hollow markers represent the (so far) non-repeating FRBs 180924 and 190523. There does not appear to be an obvious strong correlation between these quantities as a function of redshift.}
            \label{dmhost_v_redshift}
        \end{figure}

\begin{figure}
            \centering
            \includegraphics[width = \columnwidth]{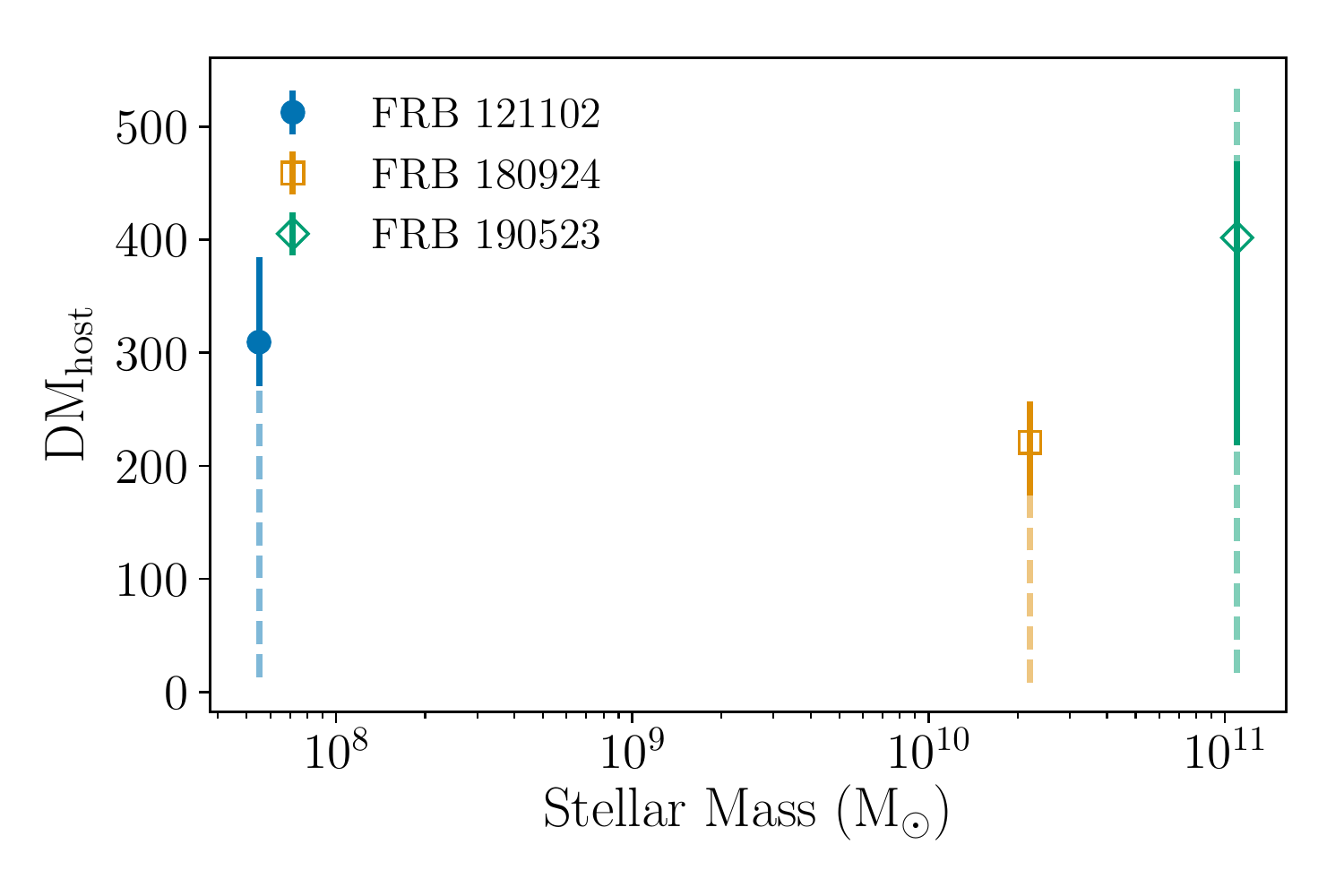}
            \caption{The host galaxy DM is plotted against the stellar mass of the host galaxy of the FRB. The solid lines represent the errors on the host galaxy DM obtained from the matter density weighted distribution while the dashed lines represent the errors obtained from the uniform weighting scheme. The solid marker represents the repeater FRB 121102, while the hollow markers represent the (so far) non-repeating FRBs 180924 and 190523. There does not appear to be an obvious strong correlation between these quantities as a function of redshift.}
            \label{dmhost_v_st_mass}
        \end{figure}

        With three FRBs now localized in redshift and to host galaxies of different types, we can begin to search for strong correlations of the host galaxy DM with other observable quantities in order to reveal information about the underlying sources of FRBs or their environments \citep[see, for example,][]{margalit_frb_src}. In Figs.~\ref{dmhost_v_redshift}  and \ref{dmhost_v_st_mass}, we plot the host galaxy DM contribution estimated using our model against the redshift and stellar mass of the host galaxy of the FRB, respectively. As we can see there do not appear to be any obvious trends, though we note that our sample size is small and the host DM might be very heavily influenced by the orientation of the galaxies and the position of the FRB within the galaxies.
        
\newpage
\section{Conclusion} \label{conclusion}
    
    In this work, we have calculated the DM contribution from the IGM using large scale cosmological dark matter simulations. We predict lower values of intergalactic DMs at low redshifts than those predicted by other works such as \citet[][]{ioka_dm}, \citet[][]{inoue_dm}, and \citet[][]{dolag_dm_igm}. Using our PDFs for the intergalactic DM contribution at different redshifts, we set upper limits on the redshifts of all the observed FRBs. The predicted redshifts are large enough that we can conclude FRBs to be extragalactic sources. Using the example of the localization of FRB 121102, we demonstrate how our intergalactic DM PDFs can be used to place constraints on the DM contribution of the host galaxy and local environment of the FRB and show that they are consistent with the DM measured using Balmer emission lines in the spectrum of the host galaxy. We also place constraints on the host galaxy DM of FRBs 180924 and 190523 which are the other two FRBs that have been localized in redshift and to a host galaxy, and look at the variation of the host galaxy DM as a function of redshift and host galaxy stellar mass.
    
    Localization of FRBs, either through instantaneous high-time-resolution transient imaging (e.g., {\it realfast}; \citealt[][]{realfast}, or the DSA110; \citealt[][]{Ravi2018}) or interferometric detection of repeaters as with FRB 121102, will allow for the determination of redshifts to FRB hosts, in which case we can use our results to place limits on the host contribution to the total dispersion measure. In the cases of other large-scale FRB surveys where localization is not immediately obtained, such as with CHIME, SUPERB at Parkes \citep[][]{SUPERB}, and UTMOST \citep{UTMOST2016,UTMOST2017}, our work provides constraints on the redshifts of FRBs. We have found that nearly all discovered FRBs so far have $z \lesssim 1$, though as new surveys are searching over a larger range of DMs, we will need to extrapolate our results either from the redshifts we have integrated over so far or from the properties of the cosmic web simulations themselves and then integrating farther in redshift space.
    
\acknowledgments

We would like to thank the MICE team, who were kind enough to provide us full access to the Onion universe data. We would also like to thank the anonymous referee for their feedback which helped to improve the manuscript.

MAM, MTL, NP, TJWL and JMC are members of the NANOGrav Physics Frontiers Center (NSF PHY-1430284). MAM and NP are supported by NSF AAG-1517003.  MAM also has additional support from NSF OIA-1458952. Part of this research was carried out at the Jet Propulsion Laboratory, California Institute of Technology, under a contract with the National Aeronautics and Space Administration.

\bibliographystyle{aasjournal}
\bibliography{bibliography.bib}

\end{document}